\documentclass[10pt]{iopart}
\usepackage{epsfig}
\usepackage{iopams}
\usepackage{amsfonts}
\usepackage{amssymb}
\usepackage{mathptmx}
\usepackage{eucal}
\usepackage{bm}

\newcommand{\Eq}[1]{Eq.~(\ref{#1})}
\newcommand{\Eqs}[1]{Eqs.~(\ref{#1})}

\begin{document}

\title{Odd-frequency superconductivity induced in topological insulators with and without hexagonal warping}

\author{A S Vasenko$^{1}$, A A Golubov$^{2,3}$, V M Silkin$^{4,5,6}$ and E V Chulkov$^{4,5,7}$}
\address{$^1$ National Research University Higher School of Economics, 101000 Moscow, Russia}
\address{$^2$ Faculty of Science and Technology and MESA$^+$ Institute for Nanotechnology,
University of Twente, 7500 AE Enschede, The Netherlands}
\address{$^3$ Moscow Institute of Physics and Technology, Dolgoprudny, 141700 Moscow, Russia}
\address{$^4$ Donostia International Physics Center (DIPC), Paseo Manuel de Lardizabal 4,
San Sebasti\'{a}n/Donostia, 20018 Basque Country, Spain}
\address{$^5$ Departamento de F\'isica de Materiales, Facultad de Ciencias Qu\'imicas,
Universidad del Pais Vasco/Euskal Herriko Unibertsitatea, Apdo. 1072, San Sebasti\'{a}n/Donostia,
20080 Basque Country, Spain}
\address{$^6$ IKERBASQUE, Basque Foundation for Science, 48011 Bilbao, Spain}
\address{$^7$ Tomsk State University, 634050 Tomsk, Russia}

\ead{avasenko@hse.ru}\vspace{10pt}

\begin{abstract}
We study the effect of the Fermi surface anisotropy on the odd-frequency spin-triplet
pairing component of the induced pair potential. We consider a superconductor/
ferromagnetic insulator (S/FI) hybrid structure formed on the 3D topological insulator (TI)
surface. In this case three ingredients insure the possibility of the odd-frequency pairing:
1) the topological surface states, 2) the induced pair potential, and 3) the magnetic moment
of a nearby ferromagnetic insulator. We take into account the strong anisotropy of 
the Dirac cone in topological insulators when the chemical potential lies well above the Dirac cone and its constant
energy contour has a snowflake shape. Within this model, we propose that the S/FI
boundary should be properly aligned with respect to the snowflake constant energy contour to
have an odd-frequency symmetry of the corresponding pairing component and to insure the
Majorana bound state at the S/FI boundary. For arbitrary orientation of the boundary the Majorana
bound state is absent. This provides a selection rule to the realization of Majorana modes in
S/FI hybrid structures, formed on the topological insulator surface.
\end{abstract}


\vspace{2pc}

\maketitle
\ioptwocol
\submitto{\JPCM}


\section{Introduction}
Three-dimensional topological insulators (3D TI) represent a recently discovered new state
of matter \cite{book1, book2, Hasan, Zhang1, Tkachov}. Their hallmark is the formation of conducting surface states with the
Dirac dispersion relation (similar to graphene), whereas the bulk states are gapped.
Recently a lot of interest has been attracted to the physics of hybrid structures involving topological
insulators and other materials, for example new electronic states have been predicted to appear in TI contacts to
ferromagnets and superconductors, such as a magnetic monopole \cite{Zhang2}
and a Majorana fermion (MF) \cite{Alicea, Leijnse, Beenakker, Beenakker2, Elliott}.
A MF has been a primary focus of many extensive studies, since it was proposed as a building block of a topological
qubit, that is robust against local decoherence \cite{Sarma}.

When the topological insulator is placed in the electrical contact with the superconductor (S), the
superconducting pair correlations penetrate into the topological state due to the
proximity effect \cite{Stanescu, Black-Schaffer2011}. Its key mechanism is the Andreev reflection process, which provides
the possibility for converting single electron states from a topological insulator to
Cooper pairs in the superconducting condensate \cite{Andreev}. Recently the S/TI proximity effect was studied by
many authors and the formation of an exotic pair potential with the so called odd-frequency spin-triplet pairing
component in the TI surface was predicted
in some particular cases, for example in the presence of the external magnetic field \cite{Asano_Tanaka, Lutchin, Snelder, Brinkman1, Brinkman2,Snelder_PRB, Burset,Huang}.

In accordance with the Pauli principle, the total wave function of a pair of fermions should be asymmetric and
can be described by the product of an orbital (or parity), spin and energy (or Matsubara frequency)
term \cite{Sigrist}. Even-frequency pairing means that a function is even in energy. If we consider singlet
(which is an odd function under spin permutation) s- or d-wave (orbitally symmetric) pairing,
the pairing wave function should be even in momentum in order for
the wave function to be antisymmetric when the pairing is even in energy. This is the so called
even-frequency spin-singlet even-parity (ESE) pairing symmetry class. For p-wave
triplet pairing the pairing wave function should be odd in the momentum. This symmetry class is referred to
as even-frequency spin-triplet odd-parity (ETO) pairing.

However, the so-called odd-frequency
pairing states when the pair amplitude is an odd function of energy can also exist.
It was first proposed by Berezinskii in the context of $^3$He \cite{Berezinskii}.
Then, the odd-frequency
spin-singlet odd-parity (OSO) and the odd-frequency spin-triplet even-parity (OTE) pairing states are allowed
by the Pauli principle. The OSO pairing is quite generally induced near the normal
metal/ superconductor interface during the formation of
the Andreev bound states \cite{Tanaka_Golubov, Tanaka_Tanuma_Golubov, Linder_Yokoyama}.
Recently it was shown, that in S/TI heterostructures another type of an odd-frequency pairing is induced in the presence of the external magnetic field,
perpendicular to the topological insulator surface - the OTE pairing state \cite{Asano_Tanaka, Lutchin, Snelder, Brinkman1,Brinkman2,Snelder_PRB, Burset, Huang}.
It was shown by Asano and Tanaka that in a one-dimensional nano-wire, proximity coupled to a topological superconductor,
the OTE pairing and the Majorana fermion are ``two sides of a same coin'' \cite{Asano_Tanaka}.
It was also argued that similar assumption is valid for two-dimensional topological surface states of a three-dimensional
topological insulator in proximity with an s-wave
superconductor \cite{Snelder, Brinkman1, Brinkman2, Snelder_PRB, Burset}.
We note here that the symmetry of the induced pair potential in S/TI structures in the diffusive case was studied in \cite{Zyuzin,Bobkova}.

A Majorana fermion is a topological state that is its own anti-particle, in striking contrast to any known
fermion so far \cite{Beenakker}. Generally, in solid state physics, electronic transport can either
be described in terms of electrons or in terms of holes. The electron and hole excitations in the superconductor play
the role of particle and antiparticle. Electrons (filled states above the chemical potential) and holes
(empty states below the chemical potential) have opposite spin and charge, but the charge difference of $2e$ can be absorbed as
a Cooper pair in the s-wave superconducting condensate.
In order for a Majorana fermion to exist, it would have
to be simultaneously half-electron and half-hole and electrically neutral. Therefore, the zero energy is a
likely place to look for a Majorana fermion. Experimentalists have already reported signatures of the Majorana
zero-energy mode, where zero-bias conductance peaks are the main features observed
in this context \cite{Deng,Mourik,Das,Finck}. Signatures of Majorana zero modes
were also observed in the Josephson effect in HgTe-based junctions \cite{Wiedenmann,Deacon},
in ferromagnetic atomic chains formed on a superconductor \cite{Nadj-Perge}, and
in a semiconductor Coulomb island in proximity with a superconductor \cite{Albrecht}.
Impressive number of theoretical studies of electron transport in different hybrid devices containing
Majorana fermions was published in recent years \cite{Fu_Kane, Tanaka, Nilsson, Fu_Majorana,
Golub, Jiang-Pekker, Badiane, Zazunov, Zazunov2, Prada, Hutzen, San-Jose, DasSarma, Rainis, Houzet, Vayrynen,
Haim, Peng}.

Majorana fermions in superconductor/ topological insulator (S/TI) hybrid structures have been first
predicted by Fu and Kane \cite{Fu_Kane} as zero energy states at the site of a vortex,
induced by the magnetic field on the surface of the topological insulator in proximity with a superconductor.
A MF was also predicted to occur if the externally applied magnetic field is replaced by the
magnetic moment of a nearby ferromagnetic insulator (FI). In the latter case,
the Majorana fermion turns out to be a one-dimensional linearly dispersing mode along the S/FI boundary, when a S/FI junction
is formed on the topological insulator surface \cite{Tanaka}.

In this work we study the interplay between topological order and superconducting correlations performing a symmetry analysis of
the induced pair potential based on the anomalous Green function and
analyze the conditions of possible realization of the Majorana mode in
a hybrid structure, where a S/FI junction is formed on the topological surface as shown in Fig.~\ref{boundary}(a) \cite{Snelder, Tanaka}.
We take into account the hexagonal warping effect, which was never considered previously in this connection,
for example in Ref.~\cite{Snelder}. Since the Majorana fermion and the odd-frequency spin-triplet even-parity (OTE) pairing
are related to each other it is important to study the effect of the hexagonal warping on the OTE pairing component realization
of the proximity effect.
In the end of the paper we make some predictions about the MF realization in the structure under consideration, based on the
symmetry arguments.


\section{Hexagonal warping}

The aforementioned studies of the proximity effect and the Majorana bound states on the surface of a 3D topological insulator were performed within
the simplest model, when the energy dispersion was described with an isotropic Dirac cone.
Within the $k \cdot p$ theory a 2$\times$2 Hamiltonian of surface states in this model to the lowest order in $k$ reads
$-\mu + v (k_x \hat{\sigma}_x + k_y \hat{\sigma}_y)$ [so called Dirac-type Hamiltonian].
Here $\mu$ is a chemical potential, $v$ is a Fermi velocity, ${\bf k} = (k_x, \;k_y)$ denotes in-plane
quasiparticle momentum, and $\hat{\sigma}_j$ are the Pauli matrices ($j = x, y, z$).
It is also possible to use a Bychkov-Rashba term \cite{Bychkov}
in the Hamiltonian of a topological insulator,
\begin{equation}
\hat{H}_0({\bf k}) = -\mu + v (k_x \hat{\sigma}_y - k_y \hat{\sigma}_x).\label{H_0}
\end{equation}
This gives rise to a different spin-momentum locking on the Fermi surface \cite{Hasan, Zhang1}.
Nevertheless, all conclusions on the excitation spectrum
are independent of whether one uses the Dirac or Bychkov-Rashba type of the Hamiltonian \cite{Linder}.
However, such isotropic forms of the Hamiltonian are only valid if the chemical potential lies near the Dirac point,
while in realistic topological insulators it usually lies well above this point, where
the Dirac cone distortion can't be any more neglected.

\begin{figure}[tb]
\epsfxsize=8.0cm\epsffile{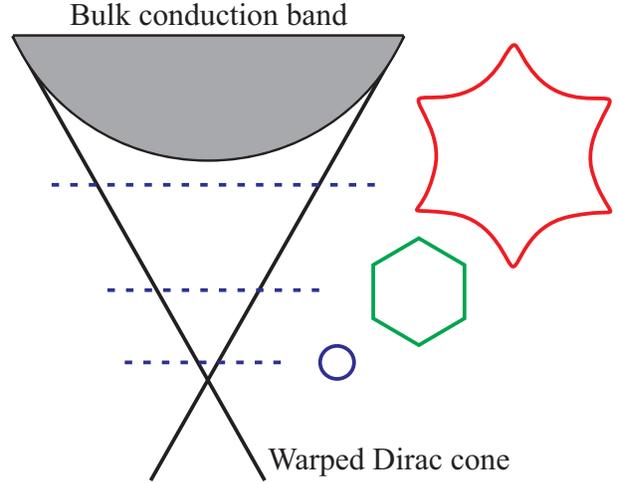} 
\caption{The hexagonal warped Dirac cone in three-dimensional topological insulators,
shown schematically. A set of constant energy contours for different chemical potential positions
(shown by dashed blue lines) is presented. Near the Dirac point
the constant energy contour is almost circular, evolving to hexagonal with increasing energy, and then to snowflake
near the bulk conduction band.}
\label{cone}
\end{figure}

To develop more realistic theoretical description of the superconducting proximity effect in
the topological insulator surface states it is important to take
into account the Dirac cone anisotropy. For example, the Fermi surface of Bi$_2$Te$_3$ topological insulator
observed by angle resolved photoemission spectroscopy (ARPES) is nearly a hexagon, having snowflake-like
shape: it has relatively sharp tips extending along six directions and curves inward in between \cite{Chen, Alpichshev, Henk}.
Moreover, the shape of constant energy contour is energy-dependent, evolving from a snowflake to a hexagon
and then to a circle near the Dirac point, see Fig.~\ref{cone}. Later same anisotropy was found in Bi$_2$Se$_3$ \cite{Kuroda, Wang},
Pb(Bi,Sb)$_2$Te$_4$\cite{Nomura} and other topological insulator materials \cite{Eremeev}.


Recently it was realized that the aforementioned simplified Hamiltonian
can be extended to higher order terms in the momentum. Namely, Fu found an unconventional hexagonal
warping term $\hat{H}_\mathrm{w}({\bf k})$ in the surface band structure, which is the counterpart of cubic
Dresselhaus spin-orbit coupling in rhombohedral structures \cite{Fu}.
The effective Hamiltonian of surface states then reads,
\begin{equation}
\hat{H}({\bf k}) = \hat{H}_0({\bf k}) + \hat{H}_\mathrm{w}({\bf k}),\label{Fu}
\end{equation}
where $\hat{H}_0({\bf k})$ is given by \Eq{H_0}, the hexagonal warping term reads
\begin{equation}
\hat{H}_\mathrm{w}({\bf k}) = \frac{\lambda}{2} (k_+^3 + k_-^3) \hat{\sigma}_z,\label{warping}
\end{equation}
$k_\pm = k_x \pm i k_y$, and $\lambda$ is the hexagonal warping strength.
The Hamiltonian in \Eq{Fu} describes perfectly the warped Dirac cone as schematically shown in Fig.~\ref{cone}.
It is the consequence of the rhombohedral crystal structure symmetry, typical to three-dimensional topological insulators.
Representative values for the material parameters $v$ and $\lambda$ can be inferred from ARPES
(see for instance \cite{Eremeev} and references therein).
We stress here that the term in \Eq{warping} is different from the trigonal warping
term in graphene \cite{Fu, trigonal}.

The warping term, \Eq{warping}, breaks the rotational symmetry of the Dirac cone. Moreover, it is an odd-parity term,
since it is odd under transformation ${\bf k} \rightarrow {\bf -k}$. It is then obvious that addition of
this term may dramatically change the physical properties of topological insulator surface states.
The consequences of warping on magnetic \cite{Fu, Jiang, Baum} and transport properties \cite{Li, Xiao}
of topological insulators have been discussed extensively in the literature.
Recently the magnetic order on a topological insulator surface with hexagonal warping and proximity-induced
superconductivity was also studied in \cite{Schon}.


\section{Model and basic equations}

We consider an s-wave superconductor/ ferromagnetic insulator (S/FI) junction formed
on the surface of a three-dimensional topological insulator in the $x$-$y$ plane, see Fig.~\ref{boundary}(a),
which was previously considered in \cite{Snelder}.
The S/FI boundary is perpendicular to the $x$-direction, so that the FI layer lies in $x>0$ half-plane, while the S layer in
$x<0$ half-plane. Since the superconducting correlations extend in the lateral direction on the
scale of superconducting coherence length, we can consider both the superconducting pair potential and the
magnetic moment in the effective ${\bf k}$-dependent Hamiltonian \cite{Snelder}, describing the TI surface states in vicinity of the S/FI boundary 
(we use ``check'' for $4 \times 4$ and ``hat'' for $2 \times 2$ matrices),
\begin{equation}
\check{H}_S({\bf k}) = \left(\begin{array}{ccc} \hat{H}({\bf k}) + M \hat{\sigma}_z & \hat{\Delta} \\
-\hat{\Delta} & -\hat{H}^* (-{\bf k}) - M \hat{\sigma}_z
\end{array}\right). \label{H_S}
\end{equation}
Here $\hat{\Delta} = i \hat{\sigma}_y \Delta$, where $\Delta$ is the superconducting pair potential,
induced in the topological insulator surface due to the proximity effect,
and $M$ is the spin-splitting (exchange) field of the ferromagnetic insulator,
which we consider to be perpendicular to the topological surface, i.e. along the $z$-axis.

In order to determine the pairing relations for the topological surface states near the S/F boundary we start with the following equation,
\begin{equation}\label{master}
\bigl [ E - \check{H}_S({\bf k}) \bigr ] \check{G} = \check{1},
\end{equation}
where $\check{G}$ is the Green's function of the topological states in vicinity of the S/FI boundary, $\check{1}$ is the unitary 4$\times$4 matrix,
and $E$ is the quasiparticle energy counted from the chemical potential.

Introducing the crystallographic angle $\theta$ which is the azimuth angle of momentum ${\bf k}$ with respect to the $x$-axis,
so that $k_x = k \cos (\theta)$, $k_y = k \sin (\theta)$, $k = |{\bf k}|$,
we can write the matrix in the left hand side of \Eq{master} in the following form,
\begin{equation}
\bigl [ E - \check{H}_S({\bf k}) \bigr ] \equiv \left(\begin{array}{ccc} \hat{h}_+ & -\hat{\Delta} \\
\hat{\Delta} & \;\;\hat{h}_-
\end{array}\right),
\end{equation}
where the matrices $\hat{h}_\pm$ are given by,
\begin{equation}
\left(\begin{array}{ccc} \epsilon_\mp \pm \mu - \lambda k^3 \cos(3\theta) & \pm i v k e^{\mp i\theta} \\
\mp i v k e^{\pm i\theta} & \epsilon_\pm \pm \mu + \lambda k^3 \cos(3\theta)
\end{array}\right),
\end{equation}
and $\epsilon_\pm = E \pm M$. The $e^{\pm i\theta}$ factors reflect the chiral odd-parity (p-wave) character of a topological insulator surface in
proximity with an s-wave superconductor.

By taking the inverse of the matrix equation (\ref{master}) we can obtain the Green's function $\check{G}$
expressed as
\begin{equation}\label{Green}
\check{G} = \left(\begin{array}{ccc} \hat{G}_\mathrm{ee} & \hat{G}_\mathrm{eh} \\
\hat{G}_\mathrm{he} & \hat{G}_\mathrm{hh}
\end{array}\right).
\end{equation}
The diagonal blocks of the $\check{G}$ matrix describe the propagation of the electrons and holes separately,
while the off-diagonal blocks describe the interaction between the electron and hole branches,
providing the mixing of the electron and hole degrees of freedom due to Andreev reflections.
To characterize superconducting pairing correlations induced in the topological surface states
we have thus to consider the the off-diagonal part of \Eq{Green},
i.e. the anomalous Green's function. Since $\hat{G}_\mathrm{eh}$ and $\hat{G}_\mathrm{he}$ are related by
complex conjugation it is sufficient to consider one of these matrices, given by the following expression,
\begin{equation}
\hat{G}_\mathrm{eh} = (\hat{\Delta} + \hat{h}_-\hat{\Delta}^{-1} \hat{h}_+)^{-1}.\label{Geh1}
\end{equation}
%


\section{Surface states without warping}\label{no_warping}

In this section we reproduce some of the results from \cite{Snelder}, rewriting all important equations in our notations for
comparison with formulae in the next section (where we take into account the hexagonal warping effect).
For now, we focus on the hybrid structure in Fig.~\ref{boundary}(a), and assume surface topological insulator states without hexagonal warping, i.e. $\lambda = 0$ in \Eq{warping}.
In this case $\hat{H}({\bf k}) = \hat{H}_0({\bf k})$ in \Eq{Fu}.

Expanding $\hat{G}_\mathrm{eh}$ in Pauli matrices (where $\hat{\sigma}_0$ is
a unitary 2$\times$2 matrix) we obtain \cite{Bergeret},
\begin{equation}
\hat{G}_\mathrm{eh} =  i \bigl(f_0 \hat{\sigma}_0 + f_x \hat{\sigma}_x + f_y \hat{\sigma}_y + f_z \hat{\sigma}_z \bigr) \hat{\sigma}_y,\label{G_an}
\end{equation}
where $f_0$ is the spin-singlet component $(\uparrow\downarrow - \downarrow\uparrow)$,
$f_x$ and $f_y$ are the combinations of equal spin triplet components, $(\uparrow\uparrow - \downarrow\downarrow)$ and
$(\uparrow\uparrow + \downarrow\downarrow)$, correspondingly, while $f_z$ is the hetero-spin triplet component,
$(\uparrow\downarrow + \downarrow\uparrow)$ \cite{Burset}. We can write these pairing wave functions explicitly as \cite{Snelder},
\numparts\label{f0}
\begin{eqnarray}
f_0 = \frac{\Delta}{Z} \left (E^2 - v^2 k^2 - B \right ),\label{f_00}
\\
f_x = \frac{2 \Delta}{Z} k v \left [\mu \sin (\theta) + i M \cos(\theta) \right ],\label{f_x0}
\\
f_y = -\frac{2 \Delta}{Z} k v \left [\mu \cos (\theta) - i M \sin (\theta) \right ],\label{f_y0}
\\
f_z = \frac{2 \Delta}{Z} E M,\label{f_z0}
\end{eqnarray}
\endnumparts
where we have used the following notations
\numparts
\begin{eqnarray}
Z = -4 k^2 v^2 (\mu^2 - M^2) - 4 E^2 M^2 + (E^2 - v^2 k^2 - B)^2,\label{Z}
\\
B = \mu^2 + \Delta^2 - M^2.\label{B}
\end{eqnarray}
\endnumparts
We note that $Z({\bf k},E)$ in \Eq{Z} is an even function in energy and momentum (even-frequency and even-parity function).
The spin-singlet component $f_0$ belongs to the even-frequency spin-singlet even-parity class (ESE). The two combinations of equal spin
components ($f_x$ and $f_y$ ) belong to the even-frequency spin-triplet odd-parity class (ETO). Finally, the hetero-spin triplet $f_z$
belongs to the odd-frequency spin-triplet even-parity class (OTE) \cite{Tanaka_Golubov}. Its presence indicates the
possibility of a Majorana fermion realization, according to recent studies \cite{Asano_Tanaka, Lutchin, Snelder, Brinkman1,Brinkman2,Snelder_PRB, Burset, Huang}.

It follows from \Eq{f_z0} that the $f_z$ pairing function appears only when three ingredients 1) the topological surface states,
2) the superconducting pair potential $\Delta$ due to the
proximity effect, and 3) the nonzero exchange field $M$ in the $z$-axis direction, are brought together (which happens
in the vicinity of the S/FI boundary in the structure under consideration).
When any of these ingredients is missing the anomalous Green's function loses the odd-frequency pairing symmetry.
It was also shown that combining these ingredients together is necessary to have the Majorana fermion realization \cite{Fu_Kane, Tanaka}.
However, the mere presence of the OTE pairing function $f_z$ still does not mean that a Majorana zero-energy mode
exists, because the zero energy mode is not yet fully localized. In Sec.~\ref{majorana} we discuss the localization problem
of the Majorana fermion when $f_z$ has pure OTE symmetry.


\section{Warped surface states}\label{yes_warping}


\begin{figure*}[tb]
\centering
\epsfxsize=12.0cm\epsffile{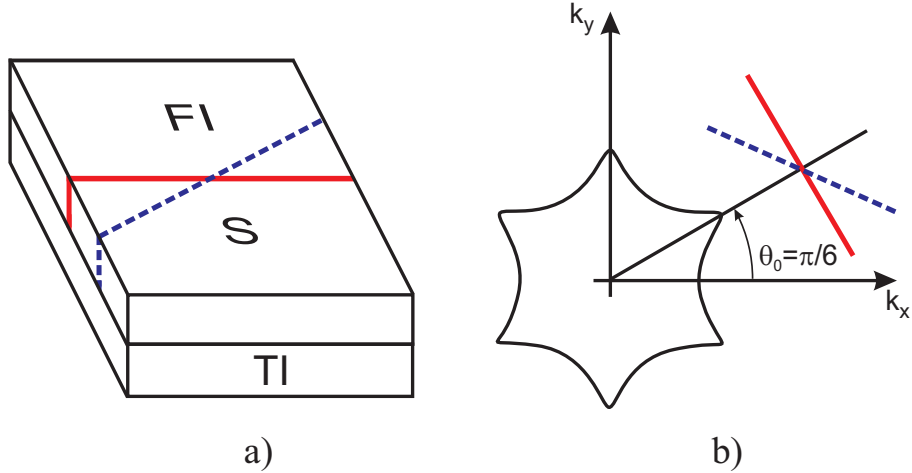} 
\caption{(a) Schematic illustration of the structure under consideration:
superconductor/ ferromagnetic insulator (S/FI) junction formed
on the surface of a three-dimensional topological insulator, TI. The $x$-axis is chosen perpendicular to the S/FI boundary.
Two possible S/FI boundaries are shown by
solid red and dashed blue lines as explained in Fig.~\ref{boundary}(b).
(b) Two possible alignments of the S/FI boundary with respect to the snowflake constant energy contour
of the warped Dirac cone. The one, shown by solid red line, is favorable for the Majorana bound state realization,
see details in the text. There are overall six favorable alignments at $\theta_n$ values, corresponding
to the six tips of the snowflake contour. The favorable alignment should be perpendicular to the line, connecting
the tip and the center of the snowflake contour. Arbitrary boundary alignments, for example the one, shown by dashed blue line, are
unfavorable and do not host the Majorana bound states.}
\label{boundary}
\end{figure*}


If we now `switch on' the hexagonal warping, so that $\lambda \neq 0$ in \Eq{warping},
the pairing wave functions in \Eq{G_an} read,
\numparts\label{f}
\begin{eqnarray}
f_0 = \frac{\Delta}{Z_+} \left (E^2 - E_S^2 - B \right ),\label{f_0}
\\
f_x = \frac{2 \Delta}{Z_+} k v \left [\mu \sin (\theta) + i M \cos(\theta) \right ],\label{f_x}
\\
f_y = -\frac{2 \Delta}{Z_+} k v \left [\mu \cos (\theta) - i M \sin (\theta) \right ],\label{f_y}
\\
f_z = \frac{2 \Delta}{Z_+} \left [ E M - \mu \lambda k^3 \cos (3 \theta) \right ].\label{f_z}
\end{eqnarray}
\endnumparts
In \Eq{f_0} $E_S$ is given by the following relation,
\begin{equation}
E_S = \sqrt{v^2 k^2 + \lambda^2 k^6 \cos^2(3\theta)}.
\end{equation}
It determines the surface band dispersion of the Hamiltonian in \Eq{Fu}, i.e. the energy dispersion
relation for the bare topological insulator surface, taking into account the hexagonal warping effect \cite{Fu},
\begin{equation}
E_{1,2} = -\mu \pm E_S.
\end{equation}
Here $E_{1,2}$ denote the energy of upper and lower band.

In \Eqs{f} the $Z_+$ function is given by,
\begin{eqnarray}
Z_\pm({\bf k},E) = - 4 k^2 v^2 (\mu^2 - M^2) + (E^2 - M^2)^2 + E_-^2 E_+^2
\nonumber
\\
- E_\mp^2 (E - M)^2 - E_\pm^2 (E + M)^2,\label{Zpm}
\end{eqnarray}
where we used the following notations,
\begin{equation}
E_{\pm} = \sqrt{v^2 k^2 + \Delta^2 + [\mu \mp \lambda k^3 \cos (3\theta)]^2}.\label{Epm}
\end{equation}
Using the equalities
\numparts
\begin{eqnarray}
Z_\pm({\bf k},E)=Z_\pm(-{\bf k},-E),
\\
Z_\pm(-{\bf k},E)=Z_\pm({\bf k},-E)=Z_\mp({\bf k},E).
\end{eqnarray}
\endnumparts
we can introduce the functions
\numparts
\begin{eqnarray}
F_\mathrm{even} = \Delta/Z_+ + \Delta/Z_-,
\\
F_\mathrm{odd} = \Delta/Z_+ - \Delta/Z_-,
\end{eqnarray}
\endnumparts
where $Z_-$ is defined in \Eq{Zpm}. It is easy to see that $F_\mathrm{even}$ is even in energy and momentum, while $F_\mathrm{odd}$ is odd in both arguments. We notice that changing the momentum sign $\bf{k} \rightarrow -\bf{k}$ is equivalent to $\theta \rightarrow \theta + \pi$
rotation.

In our spin basis the spin-triplet $f_x$ and $f_y$ pairing functions are the combinations
of equal spin components, $(\uparrow\uparrow - \downarrow\downarrow)$ and
$(\uparrow\uparrow + \downarrow\downarrow)$, correspondingly, while the function $f_z$ is the hetero-spin triplet component,
$(\uparrow\downarrow + \downarrow\uparrow)$.
Therefore only $f_z$ is providing the spin-mixing, required for the realization of an electron-hole superposition due to
Andreev reflections at the s-wave superconductor interface, which can
form a Majorana fermion under certain conditions \cite{Beenakker}.
It can be written in the following symmetrized form $f_z = f_z^- + f_z^+$, where
\numparts
\begin{eqnarray}\label{f_z2}
f_z^- = EM F_\mathrm{even} - \mu\lambda k^3 \cos(3\theta) F_\mathrm{odd},\label{fodd}
\\
f_z^+ = EM F_\mathrm{odd} - \mu\lambda k^3 \cos(3\theta)F_\mathrm{even}.\label{feven}
\end{eqnarray}
\endnumparts

If we take into account the hexagonal warping, the $f_z$ component is no more odd in energy, as in the previous section \ref{no_warping},
since the $f_z^+$ term is even in energy
for any chosen $\bf{k}$ direction, see \Eq{feven}. The $z$-component of the anomalous Green's
function became odd in energy only at the following six values of the angle $\theta$,
\begin{equation}
\theta_n = \pi/6 + \pi n/3,\label{theta}
\end{equation}
which correspond to the six tips of the snowflake constant energy contour, shown in Fig.~\ref{boundary}(b) ($n$ is an integer number).
At these values of $\theta$ the $\cos(3\theta)$ term tends to zero. In the same time $F_\mathrm{odd} = 0$ since
$E_+ = E_- = \sqrt{v^2 k^2 + \Delta^2 + \mu^2}$ in \Eq{Epm} and $Z_+ = Z_- \equiv Z$. As a result at $\theta = \theta_n$
the hexagonal warping effectively disappears and the $f_z$ pairing function is given by \Eq{f_z0} in Sec.~\ref{no_warping}.
So, at $\theta = \theta_n$ we have a possibility of the Majorana fermion realization.


\section{Majorana modes}\label{majorana}

Let us now consider six lines $\theta = \theta_n$ on the TI surface, where the topological
surface spectrum is given by the following relation [which can be obtained by diagonalization of the Hamiltonian in \Eq{H_S}],
\begin{eqnarray}
E_{1,2,3,4}= (\pm)(\mp 1)\times
\\
\sqrt{\mu^2 + \Delta^2 + v^2 k^2 + M^2 \pm 2\sqrt{\mu^2 v^2 k^2 + M^2 [\mu^2 + \Delta^2]}}.\nonumber
\end{eqnarray}
In this expression we need to take all possible combinations of $\pm$ signs to get four energy bands.
In the limit of $\mu \gg \Delta$, which is often the case, this can in good approximation be written as \cite{Snelder}
\begin{equation}
E_{1,2,3,4}= \pm (\mp) \mu \pm \sqrt{v^2 k^2 + M^2}.
\end{equation}

As was mentioned in Sec.~\ref{no_warping}, in order to have a zero-energy Majorana mode for $\theta=\theta_n$,
it needs to be localized both at the superconducting side and at the side of the
ferromagnetic insulator. At the superconducting side it happens due to the superconducting gap. At the
FI side the exchange field $M$ has to be large enough so that the chemical potential is inside the gap \cite{Snelder},
$M > \sqrt{\Delta^2 + \mu^2}$, or just $M > \mu$ if $\mu \gg \Delta$. Then the zero-energy mode is fully localized.
As was shown in \cite{Snelder} the midgap Andreev bound state became the Majorana zero mode
only when the incident electron and the Andreev-reflected hole trajectories are perpendicular to the S/FI boundary ($k_y \sim 0$).
As follows from the above arguments to insure the MF existence the S/FI boundary should be aligned perpendicular
to the line $\theta=\theta_n$.

In Fig.~\ref{boundary}(b) we show the alignment of the S/FI boundary (shown by solid red line)
which insures the Majorana zero energy mode existence
in particular case $\theta_0 = \pi/6$. Generally, the boundary should be perpendicular to the line $\theta=\theta_n$
in the topological surface plane and totally six favorable alignments are possible. The expressions for the Majorana zero
modes given in \cite{Snelder} hold in our case for $\theta=\theta_n$. In the vicinity of
$\theta=\theta_n$ the bound state energy is $E = -\Delta\sin(\theta-\theta_n)$ and goes to zero at $\theta=\theta_n$.

For $\theta \neq \theta_n$, the surface bound states dispersion relations for finite $\lambda$ become very cumbersome and we do not
present them here. But, importantly, based on the assumption on direct correspondence between the hetero-spin pure OTE component and
the Majorana fermion in 2D topological surface \cite{Snelder}, one can provide the following general symmetry-based argument.
Since the triplet component $f_z$ of the anomalous Green's function
is neither odd nor even in energy in this case [see \Eqs{f_z2}], no zero-energy Majorana bound states will be formed
for any alignment of the S/FI boundary except six aforementioned.
The example of such unfavorable alignment is shown in Fig.~\ref{boundary}(b) by a blue dashed line.
This may provide a selection rule to the realization of Majorana modes in
S/FI hybrid structures, formed on the topological insulator surface.
Recently another type of a selection rule for the MF realization was established in \cite{Annica}.


\section{Conclusion}

In conclusion, we have theoretically discussed the proximity effect in three-dimensional topological insulators with warped surface state, in the presence of a magnetic moment, perpendicular to the TI surface. For this we have considered a superconductor/ ferromagnetic insulator heterostructure, formed on the surface of a topological insulator, see Fig.~\ref{boundary}(a).

We have discussed the hetero-spin odd-frequency spin-triplet even-parity (OTE) pairing, 
induced in the warped TI surface state by proximity with a superconductor
and its relation with the Majorana fermion state. In a one-dimensional nano-wire the pure OTE pairing and the Majorana zero mode ``are
one and the same thing'' \cite{Asano_Tanaka, Lutchin}. It was also argued that same statement is valid in the two-dimensional case of
a 3D topological insulator surface \cite{Snelder}. Then the Majorana fermion realization is sensible to the orientation of the S/FI boundary with respect to the snowflake constant energy contour of the warped Dirac cone. The favorable alignment should be perpendicular to the line, connecting
the snowflake tip and the center of the snowflake contour. Arbitrary boundary alignments are unfavorable and do not host the Majorana bound states.
This may provide a selection rule to the realization of Majorana modes in
S/FI hybrid structures, formed on the topological insulator surface.


\ack{The authors thank F.S. Bergeret, P. Burset, and I.V. Tokatly for useful discussions.
The article was prepared within the framework of the Academic Fund Program at the National Research University Higher School of Economics (HSE) in 2016-2017 (grant \# 16-01-0051) and by the Russian Academic Excellence Project ``5-100''.
The research carried out by A.A. Golubov was supported by Russian Science Foundation project \# 15-12-30030.}

\end{document}